\documentclass[10pt, conference]{IEEEtran}
\IEEEoverridecommandlockouts

\usepackage{cite}
\usepackage{booktabs}
\usepackage{amsmath,amssymb,amsfonts}
\usepackage{algorithmic}
\usepackage{graphicx}
\usepackage{textcomp}
\usepackage{xcolor}
\usepackage{todonotes}
\usepackage{tikz}
\usepackage[most]{tcolorbox}
\usepackage{hyperref}
\usepackage{pifont}
\usepackage{xcolor}
\usepackage[utf8]{inputenc}
\usepackage{authblk}
\usepackage{hyperref}

\usepackage[caption=false,font=footnotesize]{subfig}

\definecolor{light-gray}{gray}{0.95}
\newcommand{\code}[1]{\colorbox{light-gray}{\texttt{#1}}}
\newcommand{\namemethod}{TreeRanker}
\newcommand{\namebaseline}{Beam@All}

\newcommand{\smartparagraph}[1]{\vspace{.05in}\noindent\textbf{#1}}
\newcommand*\circled[1]{%
  \tikz[baseline=(char.base)]{
    \node[shape=circle, draw=black, fill=black, text=white, inner sep=1pt] (char) {\sffamily\bfseries\footnotesize #1};}}

\begin{document}

\title{\namemethod{}: Fast and Model-agnostic Ranking System for Code Suggestions in IDEs}
\renewcommand\Authands{, } 

\author[1,2]{\href{mailto:daniele.cipollone@jetbrains.com}{Daniele Cipollone}}
\author[1]{\href{mailto:egor.bogomolov@jetbrains.com}{Egor Bogomolov}}
\author[2]{\href{mailto:Arie.vanDeursen@tudelft.nl}{Arie van Deursen}}
\author[2]{\href{mailto:M.Izadi@tudelft.nl}{Maliheh Izadi}}

\affil[1]{JetBrains, Amsterdam, Netherlands}
\affil[2]{Delft University of Technology, Delft, Netherlands}

\maketitle

\begin{abstract}
Token-level code completion is one of the most critical features in modern Integrated Development Environments (IDEs). It assists developers by suggesting relevant identifiers and APIs during coding. While completions are typically derived from static analysis, their usefulness depends heavily on how they are ranked, as correct predictions buried deep in the list are rarely seen by users. Most current systems rely on hand-crafted heuristics or lightweight machine learning models trained on user logs, which can be further improved to capture context information and generalize across projects and coding styles.
In this work, we propose a new scoring approach to ranking static completions using language models in a lightweight and model-agnostic way. Our method organizes all valid completions into a prefix tree and performs a single greedy decoding pass to collect token-level scores across the tree. This enables a precise token-aware ranking without needing beam search, prompt engineering, or model adaptations. The approach is fast, architecture-agnostic, and compatible with already deployed models for code completion.
These findings highlight a practical and effective pathway for integrating language models into already existing tools within IDEs, and ultimately providing smarter and more responsive developer assistance.


\end{abstract}

\begin{IEEEkeywords}
Ranking, Code Suggestions, Language Models, Integrated Development Environments, IDEs
\end{IEEEkeywords}

\section{Introduction}
Code completion features are among the most frequently used functionalities in modern Integrated Development Environments (IDEs)~\cite{murphy_how_2006, amann_study_2016,izadi2024language}. Despite the growing capabilities of Large Language Models (LLMs) to generate large code snippets~\cite{guo_deepseek-coder_2024}, developers continue to rely heavily on token-level completion~\cite{wang2023practitioners,izadi2022codefill} due to its speed, precision, and seamless integration within the coding workflow. It assists developers by suggesting relevant identifiers, functions, or API elements during typing, helping reduce effort, minimize errors, and navigate large codebases more effectively. The precision and responsiveness of the completion systems play a critical role in improving developer productivity and code quality.

Most IDEs generate completion candidates using static analysis~\cite{pei_better_2023, shrivastava_repository-level_2023}, which efficiently extracts valid suggestions from the program state. However, the effectiveness of these systems depends not only on which completions are retrieved but also on how they are ranked. For example, JetBrains IDEs rely on a proprietary machine learning model trained on anonymized usage logs to sort static completions~\cite{bibaev_all_2022}. Although these models are lightweight and efficient, they rely on hand-crafted features and are largely blind to the broader semantic context of the source file. As a result, ranking decisions are often based on metadata such as selection history, symbol frequency, or usage patterns, \emph{rather than the actual semantics of the surrounding code.}

LLMs offer a new opportunity to enhance this ranking step. While primarily used for generative tasks, LLMs also expose rich token-level probability distributions that can be repurposed to score and rank static completion candidates. However, their application in this context remains underexplored, particularly under the runtime constraints typical of local development tools.

We propose \namemethod{}, a lightweight decoding-time sorting mechanism that uses an LLM to rank a predefined set of static completions. These completions are organized into a \emph{completion tree}, where each path corresponds to a valid token sequence that represents an identifier. As the model performs greedy decoding, we collect token-level probabilities not only for the decoded path but also for all valid alternatives reachable in the tree. This allows us to construct a fine-grained ranking signal without requiring beam search~\cite{freitag_beam_2017, huang_when_2017} or prompt augmentation~\cite{fan_survey_2024, ahmed_automatic_2024}.

\namemethod{} is fast, model-agnostic, and non-intrusive. It does not require modifying the model or retraining and operates using a single greedy decoding loop. A lightweight masking strategy is used to restrict generation to valid continuations, although we find that the primary performance gains arise from the structured score collection itself, not from hard constraints. \namemethod{} is designed to work in perfect synergy with code generation models already in use within IDEs~\cite{semenkin_full_2024}.

Unlike approaches that improve code completion through prompt augmentation~\cite{reddy_first_2024}, retrieval-augmented generation~\cite{chen_apigen_2024}, or multi-step ranking~\cite{qin_large_2024}, our method operates entirely within the model’s native decoding process. It avoids the need for large prompts or multiple passes, making it highly compatible with already implemented single-pass inference pipelines. This design enables effective ranking even with compact language models. We demonstrate that our approach performs competitively using models as small as $135$ million parameters. This makes our solution viable for integration into local development environments without sacrificing latency or ranking quality.
We evaluated our method on two benchmarks that span both language diversity and the scope of completion. The first, \emph{DotPrompts}~\cite{agrawal_monitor-guided_2023}, focuses on Java dereference completions and includes a mix of global APIs and locally defined elements, providing broad coverage of typical IDE completion scenarios. The second, \emph{StartingPoints}, is a curated subset of the Long Code Arena benchmark~\cite{bogomolov_long_2024}, specifically designed to target completions involving identifiers defined within the local scope of large Python projects. This refined benchmark emphasizes realistic project-specific completions and allows us to evaluate the method’s ability to resolve and rank repository-specific APIs.
\\
In summary, we investigate the following research questions:

\circled{1} Can token-level scoring from constrained greedy decoding improve the ranking of static completions compared to existing in-IDE solutions and LLM-based baselines? \circled{2} Is the method efficient enough to be used in low-latency code completion scenarios?
\\
\textbf{Our contributions are:}
\begin{itemize}
  \item We propose \textbf{\namemethod{}}, a novel LLM-based ranking method for static code completions that uses a decoding-time tree traversal to collect token-level scores.
  \item We introduce \textbf{StartingPoints}, a new dataset for evaluating completion ranking on locally defined identifiers, based on the Long Code Arena benchmark.
  \item We evaluate the method in global and project-specific completion scenarios using Java and Python datasets to show robust performance across languages and scopes.\footnote{Our code and data will be shared publicly upon acceptance.}
  \item We show that our method outperforms existing ML and heuristic ranking systems while preserving the low latency of greedy decoding.
\end{itemize}




\section{Background}
\smartparagraph{Autoregressive Decoding.}
LLMs generate text and code using \emph{autoregressive decoding}~\cite{vaswani_attention_2023}, producing one token at a time based on the previously generated sequence. 





At each step, given a context $x = (x_1, \ldots, x_t)$, the model predicts a distribution over the vocabulary $\mathcal{V}$ for the next token:

\begin{equation}
P(x_{t+1} \mid x_1, \ldots, x_t)
\label{eq:next-token}
\end{equation}

The decoding strategy defines how tokens are selected from this distribution. In \emph{greedy decoding}, the model chooses at each step the token with the highest probability, offering speed and determinism, but potentially missing globally optimal sequences. \emph{Beam search}, by contrast, explores multiple candidate sequences in parallel and retains $k$ beams by cumulative log-probability, improving exploration for a better quality result at the cost of increased latency and computational overhead~\cite{freitag_beam_2017, huang_when_2017}.

\smartparagraph{Constrained Decoding.}
\label{sec:constrained-decoding}
To improve both the efficiency and correctness of the decoding process, we employ a grammar-constrained decoding strategy based on token masking~\cite{geng_grammar-constrained_2024}. At each step, we restrict the model’s output space to only those tokens that are valid continuations from the current token position. This is implemented via a binary mask $m \in \{0, 1\}^{|\mathcal{V}|}$ over the vocabulary $\mathcal{V}$, applied to the model’s logits. The modified logits $\tilde{L}$ are defined as:

\begin{equation}
\tilde{L}_t =
\begin{cases}
L_t & \text{if } m_t = 1 \\
-\infty & \text{otherwise}
\end{cases}
\label{eq:logit_masking}
\end{equation}

This ensures that the model cannot assign probability mass to invalid continuations. In addition, we define the mask to allow only completion-ending tokens (i.e., tokens that can terminate an identifier) when the current token position corresponds to a terminal point for one or more completions. 

\smartparagraph{Static Analysis.}
To extract valid completions for identifier prediction, we rely on static program analysis, a well-established technique for reasoning about source code without execution~\cite{pei_better_2023, shrivastava_repository-level_2023}. In our setting, we focus on predicting the next identifier following a dereference operator, such as \code{.} in Java or Python. 
IDEs and language servers support incremental parsing, which allows the construction of partial Abstract Syntax Trees (ASTs) containing placeholder nodes (e.g., \code{[UNKNOWN]}) to represent incomplete expressions at the cursor position. 
Each AST node may carry information such as variable types, bindings to declarations, and visibility constraints. 
Once this set of possible completions is extracted, the model generation is constrained to this list of valid continuations. As a result, decoding is explicitly type-constrained: the model is only allowed to score completions that are both syntactically valid and semantically consistent with the program’s type context.

\section{Methodology}

\begin{figure*}[t]
  \includegraphics[width=\linewidth]{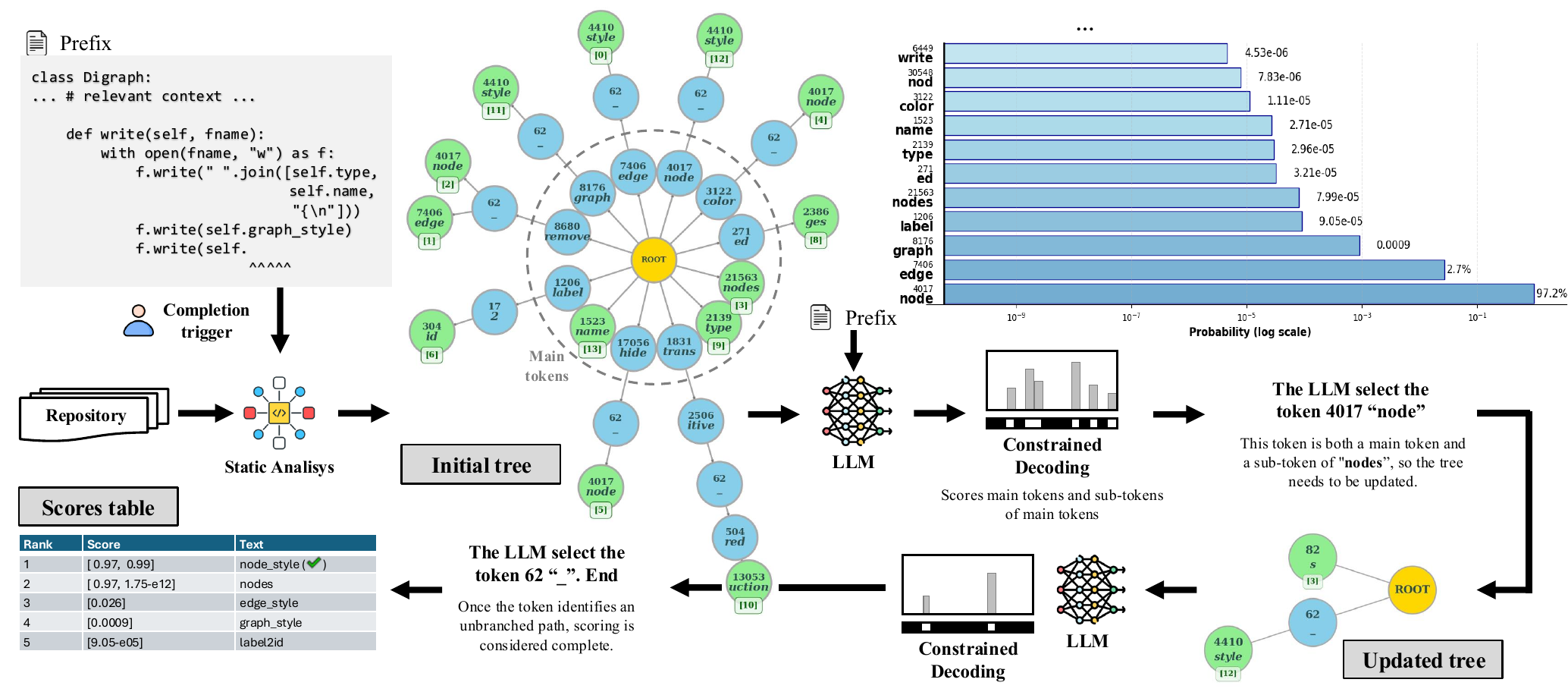}
  \caption{Visual overview of \namemethod{} scoring process over an example from StartingPoints dataset.}
  \label{fig:methodology-figures}
\end{figure*}

Our objective is to design a code completion ranking method that can enhance the quality of suggestions in real-world development environments. This imposes two key constraints: (i) the method must remain effective even when used with small language models without relying on extended context windows; (ii) the total latency of inference and ranking must fall within the strict time budgets expected for basic code completion in IDEs~\cite{kosinski2008literature, semenkin_full_2024}. With this in mind, we propose a new ranking methodology \emph{\namemethod{}}, Figure~\ref{fig:methodology-figures} illustrates an example of the ranking process. It leverages off-the-shelf language models in their standard generative setting, using the unmodified code context to score static completion candidates, and guides the process through a prefix tree constructed via greedy tokenization of candidates derived from static analysis.

\subsection{Completion Tree}
The backbone of our approach is the \emph{completion tree}.
This tree encodes the valid completions at the current cursor position and serves as a guide for efficient traversal and scoring as decoding progresses.
Let $\mathcal{V}$ denote the model's vocabulary, and let $\mathcal{C} = \{c_1, \ldots, c_n\}$ be the set of valid identifier completions. Each identifier $c_i \in \mathcal{C}$ is a string, which we tokenize using the model's greedy tokenizer $\tau$~\cite{uzan_greed_2024} to obtain a sequence of tokens:

\begin{equation}
\tau(c_i) = (t_{i,1}, \ldots, t_{i,k_i}) \in \mathcal{V}^{k_i}
\label{eq:tokenization}
\end{equation}

We organize these token sequences into a \emph{completion tree} $\mathcal{T}$, implemented as a prefix trie:

\begin{equation}
\mathcal{T} = \text{Trie}\left( \left\{ \tau(c_i) \mid c_i \in \mathcal{C} \right\} \right)
\label{eq:trie}
\end{equation}

Each node in the tree corresponds to a token prefix that forms a part of one or more valid completions. Nodes store a list of children, each connected by a token edge, and maintain a reference to the set of identifiers that share the current prefix. When a node represents the end of a full token sequence for an identifier, it is marked as a terminal node and stores a pointer to the corresponding completion.
Because identifiers are distinct and typically do not share arbitrary suffixes, they can be cleanly organized into a prefix tree, where each path from the root to a leaf corresponds to the tokenized form of a valid completion. 
A key aspect of our method relies on an \textbf{optimal decoding path optimization}: \emph{we consider that each identifier corresponds to a deterministic token sequence obtained through greedy tokenization~\cite{uzan_greed_2024}, and we only fall back to more complex processing (such as handling subtokens or tokenizer artifacts) when necessary.} This allows us to minimize the number of decoding steps while still preserving the ability to recover correct completions. Our experiments show that this fallback is required in only 12\% of cases with a marginal computational cost.

\subsection{Scoring Process}

During inference, let $x = (x_1, \ldots, x_m)$ denote the prefix context (i.e., the code before the cursor). At each step of tree traversal, we use the language model $\mathcal{M}$ to compute the next-token probabilities. Given the context $x$ and a token prefix $(t_1, \ldots, t_k)$ associated with node $v$, we obtain:

\begin{equation}
\mathcal{M}(x, t_1, \ldots, t_k) = \left\{ P(t \mid x, t_1, \ldots, t_k) \right\}_{t \in \mathcal{V}}
\label{eq:model_probs}
\end{equation}

We then collect the probabilities for each valid continuation token $t$ such that $\text{Child}(v, t) \in \mathcal{T}$. These are used to extend the token-level probability traces for each candidate identifier $c_i$.
For every identifier $c_i \in \mathcal{C}$, we maintain a sequence of probabilities corresponding to the tokens in $\tau(c_i)$ for which we have obtained model scores:

\begin{equation}
\Phi(i) = (p_1, \ldots, p_{\ell_i})
\label{eq:phi}
\end{equation}

Here, $\ell_i$ denotes the number of tokens in $\tau(c_i)$ that have been scored so far during traversal.

One of the key advantages of the completion tree structure is the ability to detect \emph{early stopping conditions} when the decoding path uniquely identifies a single valid completion. In many cases, only the first one or two tokens are sufficient to disambiguate the intended identifier (see Section \ref{sec:tree-man}). When this occurs, decoding can be stopped early, significantly reducing the number of steps required and further minimizing the overhead introduced by the ranking system.




\subsection{Tokenization details}
As discussed earlier, our method relies on an \emph{optimal decoding path optimization}, where each identifier is expected to follow a deterministic token sequence under greedy tokenization. This allows us to treat the completion tree as a fixed structure. However, tokens are not simple alphanumeric units but subword fragments, and distinct tokens may share common prefixes. This can introduce ambiguity in the decoding process that cannot be resolved by token identity alone.
For example, if \code{is} and \code{isEmpty} are treated as distinct tokens, selecting \code{is} does not exclude \code{isEmpty} as a possible continuation.
To address this, we define \emph{main tokens} as the sequence returned by greedy tokenization for each identifier. Each \emph{main token} represents the longest valid match at its position in the string. We then define \emph{subtokens} as all valid tokens that that are strict prefixes of a \emph{main token} but are not selected by the greedy tokenizer. These relationships are tokenizer-specific and allow us to build bidirectional mappings between \emph{main tokens} and their corresponding \emph{subtokens}.
At decoding time, we use these mappings in three key ways:

\textbf{(i)} We apply token-level constraints through logit masking (see Section~\ref{sec:constrained-decoding}), enabling both \emph{main tokens} and \emph{subtokens} to be treated as valid candidates at each decoding step. This improves the decoding fluency, preserving as much as possible the model distribution during generation. For example, if only \code{isEmpty} is a valid completion, we still allow the selection of \code{is} without operating directly with the tokenizer.

\textbf{(ii)} If a \emph{subtoken} is selected during decoding (e.g., \code{is}), which means that it is assigned a high probability under the model's distribution as defined in Equation~\ref{eq:model_probs}, we interpret it as a prefix of one or multiple \emph{main tokens}. In response, if it is a shared prefix of multiple \emph{main tokens}, we dynamically restructure the tree by introducing the \emph{subtoken} as a new intermediate node.
All affected completions are reassigned to this node, and their suffixes are re-tokenized (removing the decoded prefix) to generate the updated branches. If the predicted \emph{subtoken} corresponds unambiguously to a single \emph{main token}, we move the selection to the full \emph{main token}. This shortcut accelerates the decoding process, though it introduces a small deviation from the model decoding distribution.

\textbf{(iii)} Finally, once the tree is updated, we continue to decode the new node and update the probability trace $\Phi(i)$ for each candidate using the probability assigned to the selected \emph{subtoken} (as in Equation~\ref{eq:phi}). This ensures that the scoring remains consistent with the actual output of the model, while maintaining compatibility with the constrained decoding and classification strategy described in Equation~\ref{eq:rank}.

\subsection{Ranking Strategy} 
Identifiers are ranked by the extent to which their token sequence has been explored in the tree, i.e., by $\ell_i$. Among identifiers with the same $\ell_i$, we use the last available token probability as a tie-breaker. Formally:

\begin{equation}
\text{Rank}(c_i) < \text{Rank}(c_j) \iff (\ell_i, \Phi(i)[\ell_i]) > (\ell_j, \Phi(j)[\ell_j])
\label{eq:rank}
\end{equation}

This scoring scheme favors completions whose token paths align most closely with the greedy decoding trajectory of the model. At the same time, it captures the local confidence of the model at each branching point, even when a candidate diverges from the path of highest probability. By collecting probabilities not only for the selected tokens, but also for all valid continuations at each visited node, we retain a broader view of the model preferences and increase robustness to small decoding deviations.
We also experimented with alternative scoring schemes that aggregate probabilities across multiple \emph{subtokens} associated with a given \emph{main token}, but found that assigning the probability of the \emph{main token} directly yielded the best empirical results. We leave a more systematic exploration of subtoken-level score aggregation as future work.


\section{Experiment Setup}

\subsection{Benchmarks} 
To rigorously evaluate our ranking system for code completion, we use two different benchmarks. First, we employ the \textbf{DotPrompts}~\cite{agrawal_monitor-guided_2023} benchmark. 
Since this dataset includes both standard library APIs and user-defined elements, it is well suited to evaluate performance on global and local identifier prediction. 
To focus more specifically on completions involving locally defined APIs, we introduce a subset of \textbf{Long Code Arena}~\cite{bogomolov_long_2024}, a benchmark for project-wide code completion. This subset, which we refer to as \textbf{StartingPoints}, consists of Python files rich in user-defined classes and functions.

\subsubsection{DotPrompts}
provides pre-extracted completion points for Java code following dereference operations (i.e., positions after the \code{.} operator), with a clear next-token prediction objective. We extend this dataset by applying both IntelliJ IDEA and VSCode completion engines at each completion point to extract ranked lists of suggestions. We report descriptive statistics in Table~\ref{tab:dataset_stats} to characterize the overlap and diversity between IntelliJ and VSCode completions.

\begin{table}[h]
\caption{Comparison statistics between IntelliJ and VSCode}
\label{tab:dataset_stats}
\centering
\begin{tabular}{lr}
\toprule
\textbf{Metric} & \textbf{DotPrompts}\\
\midrule
\multicolumn{2}{l}{\textbf{Dataset Coverage}} \\
\midrule
Total examples                       & 7332   \\
Avg. list length (IntelliJ)          & 60.5 \\
Avg. list length (VSCode)       & 60.3  \\
Median list length (IntelliJ)        & 37     \\
Median list length (VSCode)     & 34     \\
\midrule
\multicolumn{2}{l}{\textbf{Overlap Between Systems}} \\
\midrule
Avg. Jaccard Similarity               & 0.85 \\
Identical completions                & 2859   \\
Completely different (no overlap)    & 0      \\
Significant diff. (similarity < 0.5) & 9.75\%  \\
\midrule
\multicolumn{2}{l}{\textbf{Engine-Specific Completions}} \\
\midrule
Avg. completions only in IntelliJ    & 6.7   \\
Avg. completions only in VSCode & 5.2   \\
\bottomrule
\end{tabular}
\vspace{0.5em}
\centering
\end{table}

\subsubsection{Long Code Arena}
\label{sec:lca}
was not originally designed for next-identifier prediction. To adapt it to our setting, we introduce \textbf{StartingPoints}, a filtered and annotated subset of Long Code Arena, specifically constructed to evaluate ranked completions of locally defined identifiers. We begin by analyzing each repository using the \texttt{tree-sitter}~\cite{noauthor_tree-sittertree-sitter_2025} parser to locate dereference positions, that is, all occurrences of the \code{.} operator. For each of these locations, we use the \texttt{Jedi}~\cite{noauthor_api_nodate} static analysis library, to resolve the type of the dereferenced object. To restrict the benchmark to project-local completions, we retain only those suggestions whose definitions are found within the same repository. Unlike the DotPrompts setting, we extend the dataset using only IntelliJ completions, as it was not feasible to obtain equivalent ranked suggestions from Visual Studio Code.
To ensure consistent and meaningful evaluation, we apply two key filtering steps. First, we discard examples where the ground-truth identifier is not present in the candidate list of IntelliJ, or where IntelliJ returns fewer than five valid completions, excluding boilerplate entries such as Python dunder methods (e.g., \code{\_\_del\_\_}). Second, we exclude completion points where the ground-truth identifier begins with an underscore (\code{\_}). This decision is driven by tokenization artifacts: tokenizers can treat tokens like \code{.\_} as atomic units, merging the dot and the underscore into a single token. Since our decoding setup assumes the dot is part of the static prefix, this behavior introduces inconsistencies and biases the model toward alphanumeric completions. Our inspection of tokenizer vocabularies confirmed that such merged tokens exist only for underscore-prefixed completions (e.g., \code{.\_}, \code{.\_\_}), while other common identifier prefixes (e.g.,\code{.f}, \code{.init}) are not tokenized similarly. As a result, underscore-prefixed completions are harder for the model to reach and are excluded from our evaluation. These excluded completion points represent only a minimal fraction of the dataset, and we leave a more principled handling of such cases to future work.

\begin{table}[h]
\centering
\caption{Key statistics for the DotPrompts and StartingPoints.}
\label{tab:dataset-key-stats}
\begin{tabular}{lrr}
\toprule
\textbf{Metric} & \textbf{DotPrompts} & \textbf{StartingPoints} \\
\midrule
Total examples & 7332 & 1487 \\
Unique repositories & 94 & 60 \\
Unique files & 776 & 125 \\
Avg. prefix length (lines) & 151.6 & 288.7 \\
Median prefix length (lines) & 96 & 206.0 \\
Avg. list length (IntelliJ) & 60.19 & 53.3 \\
Median list length (IntelliJ) & 37 & 39  \\
\bottomrule
\end{tabular}
\end{table}


\subsection{Metrics}
\smartparagraph{Quality Metrics.} To assess the quality of different ranking approaches, we employ a comprehensive set of evaluation metrics. We report \emph{Recall@K} for $K \in \{1, 5, 20\}$, which measures the proportion of times the correct completion appears within the top-$K$ suggestions. We also include \emph{Mean Reciprocal Rank (MRR)}, which captures the average inverse rank of the first correct suggestion. 


\emph{Among the evaluated metrics, \textbf{MRR} and \textbf{Recall@5} stand out as particularly relevant for the intended use case.} Since developers typically focus on the top few suggestions in an IDE, success is defined not only by retrieving the correct identifier but also by ranking it highly in the suggestion list.

\smartparagraph{Performance Metrics.}
We report two metrics to assess the latency of our approach: the total inference time and the \emph{ranking time}, defined as the interval between the generation of the first token and the end of the decoding process. The ranking time isolates the runtime impact of our ranking strategy, excluding the time required to produce the first token, which is highly sensitive to various factors, including hardware-specific optimizations, implementation details, and cache policies. Since optimizing this initial latency is outside the scope of our work and largely orthogonal to the proposed method, we adopt a conservative fixed estimate of 75 milliseconds~\cite{bibaev_all_2022} for the first token latency, consistent with previous measurements from ONNX Runtime~\cite{noauthor_onnx_nodate} and llama.cpp~\cite{noauthor_ggml-orgllamacpp_2025} implementations on consumer hardware.
By focusing on the time elapsed between the generation of the first token and the end of decoding, we isolate the performance impact of the ranking system and obtain a more accurate estimate of its overhead in a real IDE deployment. 
To quantify efficiency, we define the \textit{Token Efficiency Ratio}, as the ratio between the number of tokens required to represent the ground truth identifier and the number of tokens generated by the model.
Values higher than $1$ indicate more optimal decoding, while values below $1$ suggest over-generation. This metric captures the alignment between the model’s decoding path and the target identifier, and reflects the efficiency of the constrained traversal.

\subsection{Model Selection}
As previously discussed, the goal of this work is to develop methods suitable for scale deployment on consumer hardware. 
For this reason, we focus on small and extremely compact open-source models, aiming to approximate the performance of custom deployments feasible on end-user hardware.
In particular, we take inspiration from IntelliJ's production setup~\cite{semenkin_full_2024}, which imposes strict constraints on memory footprint and inference time. Their completion engine leverages a quantized LLaMA-like~\cite{touvron_llama_2023} model with approximately 100M parameters and a maximum context window of $1,536$ tokens. 

Motivated by these practical considerations, our study focuses on models ranging from 130M to 1.3B parameters. 
The selected models are: \texttt{SmolLM2-135M}~\cite{allal_smollm2_2025}, 
\texttt{SmolLM2-360M}~\cite{allal_smollm2_2025} 
\texttt{codegen-350M-multi}~\cite{nijkamp_codegen_2023}, 
and \texttt{deepseek-coder-1.3b-base}~\cite{guo_deepseek-coder_2024}. We also include the SmolLM models in our study, despite their general-purpose nature, because they represent a unique class of extremely compact models. Although not exclusively trained for code, they are among the few recent models of this scale that have been exposed to coding tasks which makes them a valuable point of comparison.


To ensure a fair comparison across different model sizes, we standardize the context window to 1920 tokens for all experiments. 
This decision is crucial given that the ranking performance is highly sensitive to the amount of contextual information provided. 
We select models with consistent tokenization of the dot operator (see Section \ref{sec:lca}); however, our methodology is general and can be adapted to any tokenization scheme by modifying the subset of target tokens.


\subsection{Baselines}
To provide concrete points of comparison for our ranking task, we include as baselines the code completion systems of the two most widely used IDEs: IntelliJ IDEA~\cite{noauthor_intellij_nodate} for both the datasets and Visual Studio Code~\cite{noauthor_visual_nodate} for DotPrompts (Java). 

\smartparagraph{IntelliJ IDEA} as described by Bibaev et al.~\cite{bibaev_all_2022}, uses a machine learning-based ranking model trained on anonymized usage logs collected from real developers. The model is implemented using \emph{CatBoost}~\cite{cao_learning_2007} with the \emph{QuerySoftMax}\cite{noauthor_ranking_nodate} loss function, which is specifically designed to prioritize the correct suggestions at the top of the completion list.

\smartparagraph{VSCode} leverages IntelliCode~\cite{silver_introducing_2018}, a plugin that augments the standard completion engine by reordering suggestions using machine learning. While the internal architecture of IntelliCode is not publicly documented in detail, it is known to personalize suggestions based on user behavior and code context, blending statistical insights with conventional ranking strategies.

In addition to IDE-based systems, we incorporate LLMs as baselines for code completion ranking, drawing inspiration from prior work (see Section \ref{sec:related_work}) that leverages beam search outputs to guide code suggestions~\cite{tabachnyk_ml-enhanced_nodate, li_toward_2021}. 
Specifically, we generate candidate completions using greedy decoding and beam search with widths of 5 and 20. 
We included a filtered variant of each beam setting, denoted as \emph{Beam@K Filtered}, where only completions present in the IntelliJ candidate list are retained. Although similar filtering could be applied using VSCode completions, our empirical analysis revealed that IntelliJ and VSCode yield comparable candidate sets.
To represent the upper limit of beam-based approaches, we introduce a powerful baseline denoted as \emph{\namebaseline{}}. It performs an exhaustive, constrained search of the completion tree, scoring each candidate by its cumulative log-probability with a length penalty. This process makes \emph{\namebaseline{} efficient and functionally equivalent to a re-ranker driven directly by the model loss function}.\footnote{Mathematical proof in future appendix.}

\subsection{Hardware Setup}
All experiments were performed on a machine equipped with a single NVIDIA RTX 4090 GPU with 24GB of memory.
Inference was executed using standard PyTorch and Hugging Face Transformers~\cite{wolf_huggingfaces_2020} and in half-precision (FP16). Given this setup, our reported inference times should be interpreted as conservative upper bounds. In practice, significant reductions in latency could be achieved through quantization and hardware-aware optimizations, particularly for deployment in real-time IDE environments.

\section{Results}

\subsection{Evaluation}
\label{sec:rq1}
\begingroup
\setlength{\tabcolsep}{3pt} 

\begin{table}
\centering
\caption{Evaluation metrics across baseline IDEs and LLM-based completion strategies on \textbf{DotPrompts} and \textbf{StartingPoints}.}
\label{tab:ranking-results}

\resizebox{\columnwidth}{!}{\begin{tabular}{llccccc cccc}
\toprule
& & \multicolumn{4}{c}{\textbf{StartingPoints}}
& \multicolumn{4}{c}{\textbf{DotPrompts}} \\
\cmidrule(lr){3-6} \cmidrule(lr){7-10}
Model & \textbf{Method}
& \textbf{MRR} & \textbf{R@1} & \textbf{R@5} & \textbf{R@20}
& \textbf{MRR} & \textbf{R@1} & \textbf{R@5} & \textbf{R@20} \\
\midrule
IDEs 
& VSCode       & - & - & - & - & 0.45 & 0.32 & 0.58 & 0.86 \\
& IntelliJ     & 0.49 & 0.33 & 0.68 & 0.90 & 0.60 & 0.46 & 0.79 & 0.92 \\
\midrule
135M 
& Greedy       & 0.56 & 0.56 &  &  & 0.62 & 0.62 &    &    \\
& Beam@5       & 0.61 & 0.57 & 0.67 &  & 0.67 & 0.63 & 0.75 &    \\
& \hspace{1em}+ Filter 
               & 0.62 & 0.58 & 0.67 &  & 0.70 & 0.66 & 0.75 &    \\
& Beam@20      & 0.62 & 0.57 & 0.68 & 0.72 & 0.68 & 0.62 & 0.76 & 0.81 \\
& \hspace{1em}+ Filter 
               & 0.65 & 0.59 & 0.72 & 0.72 & 0.73 & 0.68 & 0.80 & 0.81 \\
& Beam@All     & 0.72 & 0.60 & 0.87 & 0.98 & 0.76 & 0.66 & 0.89 & 0.98 \\
\midrule
& \textbf{\namemethod{}}
               & 0.73 & 0.62 & 0.88 & 0.97 & 0.76 & 0.67 & 0.87 & 0.97 \\
\midrule
350M
& Greedy       & 0.64 & 0.64 &  &  & 0.72 & 0.72 &    &    \\
& Beam@5       & 0.69 & 0.65 & 0.75 &  & 0.77 & 0.72 & 0.83 &    \\
& \hspace{1em}+ Filter 
               & 0.71 & 0.67 & 0.75 &  & 0.79 & 0.76 & 0.83 &    \\
& Beam@20      & 0.69 & 0.65 & 0.75 & 0.78 & 0.78 & 0.72 & 0.84 & 0.88 \\
& \hspace{1em}+ Filter 
               & 0.72 & 0.68 & 0.78 & 0.78 & 0.82 & 0.77 & 0.87 & 0.88 \\
& Beam@All     & 0.76 & 0.67 & 0.88 & 0.97 & 0.84 & 0.76 & 0.94 & 0.99 \\
\midrule
& \textbf{\namemethod{}}
               & 0.78 & 0.70 & 0.89 & 0.95 & 0.83 & 0.77 & 0.91 & 0.97 \\
\midrule
360M 
& Greedy       & 0.63 & 0.63 &  &  & 0.69 & 0.69 &    &    \\
& Beam@5       & 0.67 & 0.64 & 0.72 &  & 0.75 & 0.70 & 0.81 &    \\
& \hspace{1em}+ Filter 
               & 0.69 & 0.65 & 0.72 &  & 0.77 & 0.73 & 0.81 &    \\
& Beam@20      & 0.68 & 0.64 & 0.73 & 0.77 & 0.75 & 0.70 & 0.82 & 0.86 \\
& \hspace{1em}+ Filter 
               & 0.71 & 0.66 & 0.76 & 0.77 & 0.79 & 0.75 & 0.85 & 0.86 \\
& Beam@All     & 0.77 & 0.68 & 0.89 & 0.98 & 0.83 & 0.75 & 0.94 & 0.99 \\
\midrule
& \textbf{\namemethod{}}
               & 0.78 & 0.69 & 0.90 & 0.98 & 0.82 & 0.75 & 0.90 & 0.97 \\
\midrule
1.3B 
& Greedy       & 0.71 & 0.71 &  &  & 0.78 & 0.78 &    &    \\
& Beam@5       & 0.75 & 0.72 & 0.79 &  & 0.82 & 0.78 & 0.88 &    \\
& \hspace{1em}+ Filter 
               & 0.76 & 0.74 & 0.79 &  & 0.85 & 0.82 & 0.88 &    \\
& Beam@20      & 0.75 & 0.72 & 0.79 & 0.82 & 0.83 & 0.78 & 0.88 & 0.91 \\
& \hspace{1em}+ Filter 
               & 0.78 & 0.75 & 0.82 & 0.82 & 0.86 & 0.83 & 0.91 & 0.91 \\
& Beam@All     & 0.83 & 0.75 & 0.93 & 0.99 & 0.88 & 0.80 & 0.96 & 0.99 \\
\midrule
& \textbf{\namemethod{}}
               & 0.84 & 0.78 & 0.92 & 0.99 & 0.88 & 0.83 & 0.94 & 0.99 \\
\bottomrule
\end{tabular}}
\end{table}

To evaluate the effectiveness of our approach, we compare \namemethod{} to standard IDE completion engines and LLM-based decoding strategies across a range of models. The results are reported in Table~\ref{tab:ranking-results}.
Across all model sizes and both datasets, \namemethod{} consistently delivers top-tier ranking performance. Outperforms IntelliJ, VSCode, and base beam decoding variants in all retrieval metrics. For instance, on \textit{DotPrompts}, \namemethod{} on SmolLM2-135M improves the mean reciprocal rank (MRR) by up to 16 points and Recall@5 by 8 points compared to IntelliJ. 

Most notably, \namemethod{} matches the performance of the strongest baseline, \emph{Beam@All}, which computes a full beam score for each candidate. The two methods yield nearly identical results (within a few percentage points) across MRR, R@5, and R@20 on both datasets and across all model scales. \emph{These results demonstrate that \namemethod{} achieves ranking quality equivalent to a full scoring pass over all candidates.}

A breakdown by model scale reveals an expected trend: larger models (e.g., DeepSeek-Coder~1.3B) obtain the best absolute performance. However, smaller models such as SmolLM2-135M still benefit meaningfully from \namemethod{} and achieve competitive results in both datasets.
The \textit{StartingPoints} dataset presents an additional challenge due to its exclusive focus on identifiers defined within the project scope. Despite the lack of visibility on local implementations, \namemethod{} maintains consistent improvements over baselines.

Beam@5-20 offers reasonable performance in Recall@1, particularly with larger models. However, when evaluating ranking quality with more informative metrics such as MRR and Recall@5, \namemethod{} clearly outperforms such metrics. Moreover, increasing the beam width from $5$ to $20$ yields only marginal gains, and some models exhibit performance saturation. 
In contrast, \namemethod{}’s improvements reflect its ability to score and prioritize completions more accurately on the entire suggestion list.


\begin{table}[t]
\centering
\caption{Performance on identifiers not present in the file prefix}
\label{tab:first_time_identifiers}
\resizebox{\columnwidth}{!}{\begin{tabular}{llcccc cccc}
\toprule
& & \multicolumn{4}{c}{\textbf{StartingPoints}} & \multicolumn{4}{c}{\textbf{DotPrompts}} \\
\cmidrule(lr){3-6} \cmidrule(lr){7-10}
\textbf{Model} & \textbf{Method} 
& \textbf{MRR} & \textbf{R@1} & \textbf{R@5} & \textbf{R@20}
& \textbf{MRR} & \textbf{R@1} & \textbf{R@5} & \textbf{R@20} \\
\midrule
IDEs 
& IntelliJ              & 0.26 & 0.10 & 0.40 & 0.92 & 0.50 & 0.37 & 0.66 & 0.89 \\
\midrule
135M 
& Beam@20+F      & 0.20 & 0.18 & 0.23 & 0.23 & 0.49 & 0.42 & 0.59 & 0.60 \\
& \namebaseline{}      & 0.45 & 0.28 & 0.66 & 0.95 & 0.59 & 0.43 & 0.79 & 0.95\\
& \textbf{\namemethod{}}         & 0.47 & 0.30 & 0.71 & 0.94 & 0.58 & 0.44 & 0.76 & 0.94 \\
\midrule
350M 
& Beam@20+F      & 0.28 & 0.24 & 0.32 & 0.32 & 0.64 & 0.57 & 0.73 & 0.74 \\
& \namebaseline{}      & 0.47 & 0.30 & 0.68 & 0.92 & 0.72 & 0.59 & 0.87 & 0.97\\
& \textbf{\namemethod{}}         & 0.49 & 0.33 & 0.69 & 0.88 & 0.69 & 0.58 & 0.83 & 0.96 \\
\midrule
360M
& Beam@20+F      & 0.25 & 0.21 & 0.29 & 0.29 & 0.60 & 0.53 & 0.68 & 0.69 \\
& \namebaseline{}      & 0.48 & 0.32 & 0.68 & 0.94 & 0.70 & 0.57 & 0.87 & 0.97\\
& \textbf{\namemethod{}}         & 0.49 & 0.33 & 0.72 & 0.96 & 0.67 & 0.55 & 0.82 & 0.95 \\
\midrule
1.3B 
& Beam@20+F      & 0.36 & 0.32 & 0.39 & 0.40 & 0.73 & 0.67 & 0.81 & 0.81 \\
& \namebaseline{}      & 0.58 & 0.43 & 0.79 & 0.98 & 0.78 & 0.67 & 0.92 & 0.98\\
& \textbf{\namemethod{}}         & 0.59 & 0.45 & 0.78 & 0.98 & 0.77 & 0.68 & 0.89 & 0.98 \\
\bottomrule
\end{tabular}}
\end{table}

\subsection{Evaluation on Unseen Identifiers}
We further investigate model performance in a more challenging setting and focus on completions where the correct identifier is not present in the input prefix. This subset of $2,676$ samples for \textit{DotPromts} and $340$ samples for \textit{StartingPoints} captures a common scenario in which the model must choose among static completions that are unseen in the context window, such as when accessing a class member or method for the first time. The results for this setting are reported in Table~\ref{tab:first_time_identifiers}.
As expected, all LLM-based methods exhibit a notable drop in Recall@1 compared to the full benchmark. This confirms that models rely heavily on the provided prefix to confidently resolve completions. This effect is consistent across model sizes and decoding strategies, including \namemethod{}. 

Even with constrained decoding, Recall@1 remains lower than in the full setting, \emph{supporting the intuition that filtering invalid completions is necessary but not sufficient for accurate top-rank selection.}
However, despite this drop in rank one, \namemethod{} continues to maintain high MRR and Recall@5 across all configurations. This indicates that the scoring strategy still effectively retrieves the correct completions from the model, even when full identifier resolution is not achievable from the context alone. These gains are particularly relevant in practical scenarios, where developers often select completions from among the top few entries during first-time exploration.
The trend holds across both benchmarks.

\textbf{Answer RQ \circled{1}:} It consistently matches or outperforms traditional IDE systems and standard LLM baselines across all model sizes and benchmarks. 
\namemethod{} achieves ranking quality equivalent to a full scoring pass over all candidates. Moreover, it maintains strong MRR and Recall@5 even in the challenging case where the target identifier is not present in the prefix.

\subsection{Performance Evaluation}
\label{sec:time}
\begin{table}
\centering
\caption{Average generation and ranking overhead times (seconds).}
\label{tab:timing_summary}
\resizebox{\columnwidth}{!}{\begin{tabular}{llcc|r}
\toprule
\textbf{Model} & \textbf{Method} & \textbf{Total Time} & \textbf{Ranking Time} & \textbf{Ranking}  \\
\cmidrule(lr){1-4}
\multicolumn{3}{l}{\textbf{StartingPoints}} & &\textbf{Speedup}\\
\midrule
SmolLM2-135M       & Greedy        & 0.421 ± 0.005  & 0.101 ± 0.001 & 1.53x \\
                   & Beam@5        & 0.486 ± 0.008  & 0.161 ± 0.002 & 2.44x  \\
                   & Beam@20       & 0.743 ± 0.008  & 0.418 ± 0.002 & 6.34x  \\
                   & Beam@All      & 1.652 ± 0.083  & 1.615 ± 0.082 & 24.47x  \\
                   & \textbf{\namemethod{}}      & \textbf{0.392 ± 0.011}  & \textbf{0.066 ± 0.003} & $\sim$ \\
\midrule
CodeGen-350M       & Greedy        & 0.602 ± 0.005  & 0.240 ± 0.002 & 1.60x  \\
                   & Beam@5        & 0.755 ± 0.009  & 0.426 ± 0.004 & 2.84x  \\
                   & Beam@20       & 1.915 ± 0.018  & 0.483 ± 0.016 & 3.22x  \\
                   & Beam@All      & 1.457 ± 0.052  & 1.426 ± 0.051 & 9.51x  \\
                   & \textbf{\namemethod{}}      & \textbf{0.522 ± 0.011}  & \textbf{0.150 ± 0.003}  & $\sim$\\

\midrule
SmolLM2-360M       & Greedy        & 0.494 ± 0.005  & 0.157 ± 0.001 & 1.57x  \\
                   & Beam@5        & 0.598 ± 0.007 & 0.258 ± 0.002 & 2.58x  \\
                   & Beam@20       & 1.201 ± 0.007 & 0.797 ± 0.002 & 7.97x  \\
                   & Beam@All      & 1.816 ± 0.110  & 1.777 ± 0.108 & 17.77x  \\
                   & \textbf{\namemethod{}}      & \textbf{0.446 ± 0.013}  & \textbf{0.100 ± 0.003}  & $\sim$\\

\midrule
DeepSeek-Coder 1.3B& Greedy        & 0.374 ± 0.003  & 0.139 ± 0.001 & 1.55x  \\
(w. Flash Attention)& Beam@5        & 0.492 ± 0.004  & 0.124 ± 0.002 & 1.38x  \\
                   & Beam@20       & 1.120 ± 0.008  & 0.220 ± 0.002 & 2.45x  \\
                   & Beam@All      & 1.583 ± 0.073  & 1.535 ± 0.072  & 17.06x \\
                   & \textbf{\namemethod{}}      & \textbf{0.328 ± 0.007}  & \textbf{0.090 ± 0.003}  & $\sim$\\
\midrule
\multicolumn{4}{l}{\textbf{DotPrompts}} \\
\midrule
SmolLM2-135M       & Greedy        & 0.398 ± 0.008  & 0.070 ± 0.002 & 1.27x   \\
                   & Beam@5        & 0.443 ± 0.006  & 0.110 ± 0.002 & 2.00x   \\
                   & Beam@20       & 0.627 ± 0.006 & 0.295 ± 0.002 & 5.36x   \\
                   & Beam@All      & 1.682 ± 0.028  & 1.644 ± 0.027 & 29.9x   \\
                   & \textbf{\namemethod{}}      & \textbf{0.330 ± 0.023} & \textbf{0.055 ± 0.005} & $\sim$\\
\midrule
CodeGen-350M       & Greedy        & 0.505 ± 0.095  & 0.126 ± 0.018 & 1.30x   \\
                   & Beam@5        & 0.642 ± 0.009 & 0.290 ± 0.004 & 2.99x   \\
                   & Beam@20       & 1.448 ± 0.015 & 0.499 ± 0.011 & 5.14x   \\
                   & Beam@All      & 1.604 ± 0.022  & 1.562 ± 0.022 & 16.98x   \\
                   & \textbf{\namemethod{}}      & \textbf{0.392 ± 0.009} & \textbf{0.097 ± 0.002} & $\sim$\\

\midrule
SmolLM2-360M       & Greedy        & 0.430 ± 0.008  & 0.092 ± 0.001 & 1.21x   \\
                   & Beam@5        & 0.520 ± 0.014 & 0.181 ± 0.004 & 2.38x   \\
                   & Beam@20       & 0.952 ± 0.007 & 0.571 ± 0.002 & 7.51x   \\
                   & Beam@All      & 1.463 ± 0.030  & 1.432 ± 0.030 & 18.84x   \\
                   & \textbf{\namemethod{}}      & \textbf{0.421 ± 0.008} & \textbf{0.076 ± 0.002} & $\sim$\\

\midrule
DeepSeek-Coder 1.3B& Greedy        & 0.318 ± 0.014  & 0.080 ± 0.004 & 1.31x   \\
(w. Flash Attention)& Beam@5        & 0.400 ± 0.005 & 0.076 ± 0.001 & 1.25x   \\
                   & Beam@20       & 0.850 ± 0.005 & 0.131 ± 0.001 & 2.15x   \\
                   & Beam@All      & 1.835 ± 0.024  & 1.794 ± 0.023 & 29.41x   \\
                   & \textbf{\namemethod{}}      & \textbf{0.257 ± 0.005} & \textbf{0.061 ± 0.002} & $\sim$\\
\bottomrule
\end{tabular}}
\end{table}

To assess the practicality of our approach in real-world settings, we measure the average inference time of each model and decoding strategy, separating the total decoding time from the time specifically attributable to the ranking mechanism. The results are reported in Table~\ref{tab:timing_summary} for both benchmarks.

To robustly assess generation latency across configurations, we conducted five independent runs for each experimental setting, collecting per-sample inference data. For each input, we recorded both the total generation time and the \textbf{ranking time}, computed as the duration from the generation of the first token to the completion of the decoding process. As discussed previously, this excludes the initial token latency, for which we adopt a conservative upper bound estimate of $75$ milliseconds based on previous interactive completion benchmarks~\cite{bibaev_all_2022}.

We report the mean generation and ranking times across all runs, along with associated confidence intervals (CIs) to quantify variability and measurement precision. CIs were computed using Student’s t-distribution~\cite{student_probable_1908}. Given the large number of completion points, $7,332$ for \textit{DotPrompts} and $1,487$ for \textit{LCA-StartingPoints}, we compute per-sample statistics by first averaging across the five runs at each fixed input position, then calculating the standard error and confidence interval for that position. The final reported values are obtained by averaging these per-sample means and confidence intervals across the dataset. Compared to computing a global confidence interval over per-run averages, this approach better reflects the stability of runtime behavior.
The performance evaluation indicates that \namemethod{} remains well within interactive latency bounds. In DotPrompts, for example, SmolLM2-135M completes the ranking in $55 \pm 5$ milliseconds. Adding the conservative 75ms upper bound for first token generation time, the total latency remains safely below the commonly accepted threshold for real-time task~\cite{kosinski2008literature, semenkin_full_2024}. On average, this performance is even faster than the model's native greedy decoding, thanks to the possibility of early stopping during generation. \namemethod{} thus achieves a favorable trade-off, delivering improvements in ranking quality (as shown in Section~\ref{sec:rq1}) while maintaining response times compatible with deployment in interactive environments. 

\emph{\namemethod{} achieves ranking quality on par with computationally expensive reranking approaches while delivering up to \textbf{30× speedup in inference time}, combining high accuracy with substantial efficiency gains.}

\begin{figure}
  \centering
  \includegraphics[width=\columnwidth]{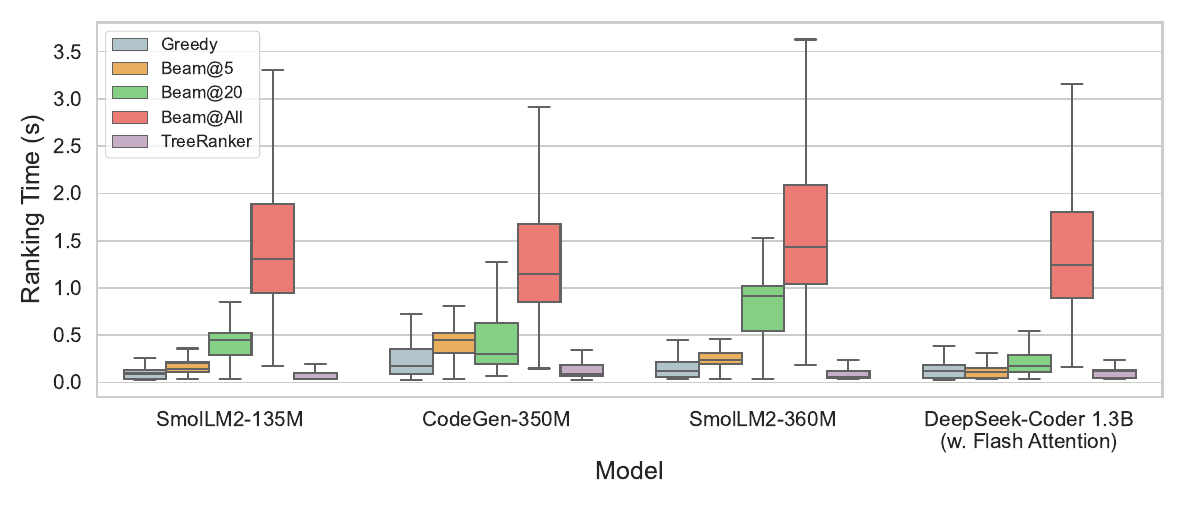}
    \includegraphics[width=\columnwidth]{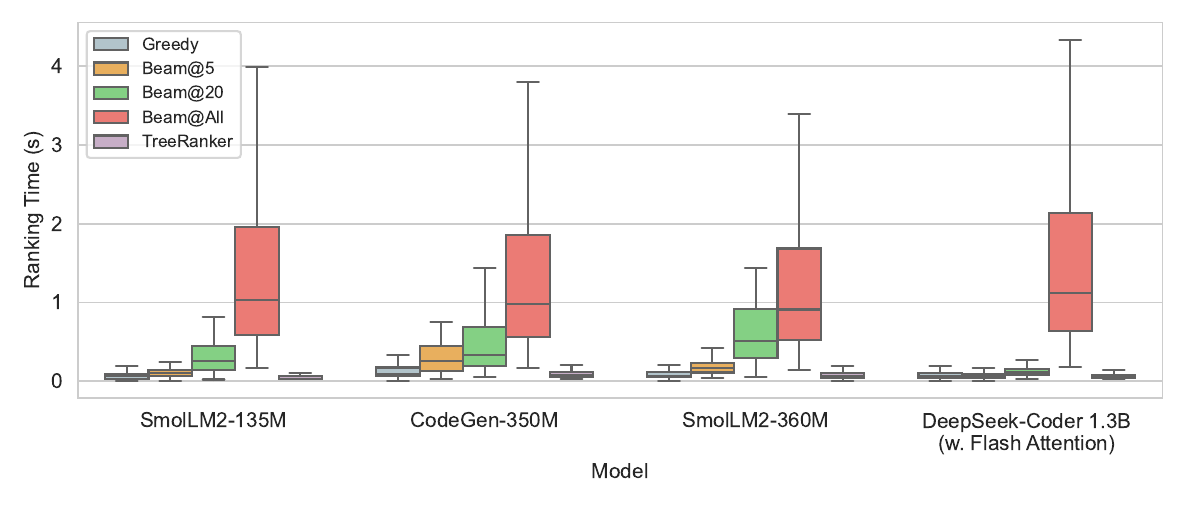}
  \caption{Distribution of ranking times across models and decoding strategies on StartingPoints(1st) and Dotprompt(2nd). \namemethod{} achieves stable, low-latency performance while maintaining high ranking perfromances.}
  \label{fig:ranking_time_boxplot}
\end{figure}

To further illustrate the distribution of the ranking times between individual inputs, Figure~\ref{fig:ranking_time_boxplot} presents a boxplot for each model and decoding method. Although Table~\ref{tab:timing_summary} summarizes overall performance using average values and confidence intervals across the dataset, this figure shows how the ranking times vary between different completion points. \namemethod{} consistently exhibits low median latency and narrow interquartile ranges, comparable to greedy decoding, and substantially more stable than beam-based approaches. 
Moreover, the figure showcases the core limitation of reranking techniques with linear complexity in the number of candidates: they introduce excessive variance in latency, which becomes a major issue in human-interactive scenarios where consistency is crucial.

\subsection{Statistics on Tree Manipulation}
\label{sec:tree-man}
To better understand the internal dynamics of our decoding strategy, we analyze structural statistics related to tree traversal using the DeepSeek-Coder 1.3B model. Similar patterns were observed across all models. As shown in Table~\ref{tab:dataset-key-stats-inference}, most completions terminate early, with $77\%$ in \textit{DotPrompts} and $89\%$ in \textit{StartingPoints} completing before fully generating the identifier. This early exit behavior indicates that the model is often able to narrow the candidate set quickly, reducing the number of decoding steps and saving computational resources.

\begin{table}[h]
\centering
\caption{Statistics on inference with \namemethod{}}
\label{tab:dataset-key-stats-inference}
\begin{tabular}{lrr}
\toprule
\textbf{Metric} & \textbf{DotPrompts} & \textbf{StartingPoints} \\
\midrule
\multicolumn{3}{l}{\textbf{Dataset stats}} \\
\midrule
Total examples                              & 7332   & 1487 \\
Avg. ground truth length (tokens)      & 3.07   & 4.47 \\
Median ground truth length (tokens)      & 3   & 4 \\
\midrule
\multicolumn{3}{l}{\textbf{\namemethod{} Features}} \\
\midrule
Early completion                            & 77\%   & 89\% \\
Gen. new tree sub-branch                    & 13\%   & 9\% \\
Main Tokens push                            & 1.4\%  & 1.3\% \\
\midrule
\multicolumn{3}{l}{\textbf{Generation}} \\
\midrule
\textbf{Avg. gen. tokens}                            & \textbf{1.86}   & \textbf{2.15} \\
Std. gen. tokens                            & 0.84   & 1.77 \\
Single forward pass                         & 35\%   & 57\% \\
Within two forward passes                  & 84\%   & 68\% \\
Token Efficiency Ratio $\uparrow$           & 1.85   & 2.77\\
\bottomrule
\end{tabular}
\end{table}

Cases where the tree must be modified to resolve ambiguities between overlapping token prefixes are rare. These cases account for just 13\% of completions in \textit{DotPrompts} and 9\% in \textit{StartingPoints}, confirming that most decoding paths align well with the optimistic tree structure built from greedy tokenization. Additionally, main token pushes are observed in only about 1.3\% to 1.4\% of completions. These low rates reflect the stability and structural alignment of the decoding process. Overall, the method proves to be highly efficient in practice, with average Token Efficiency Ratios of 1.85 and 2.77 on the two datasets, showing that most completions are ranked correctly with significantly fewer decoding steps than the full tokenized length of the target identifier.

\textbf{Answer RQ \circled{2}:} \namemethod{} is capable of ranking full candidate lists with high precision by extracting full knowledge of the model using only 1.86 and 2.15 tokens on average per completion. It matches the ranking quality of computationally expensive reranking approaches while achieving up to a 30× speedup in inference time. These properties make \namemethod{} well suited for interactive code completion scenarios.

\subsection{Ablation Study}
Constrained decoding is not introduced as a core contribution, but rather as an optimization to reduce the number of decoding steps during prefix tree traversal. Crucially, it is not required for the ranking itself, which is solely based on the token probabilities of the model. As shown in Table~\ref{tab:ablation_constraint_fixed}, performance in MRR and recall metrics remains effectively unchanged with / and without constraints. Figure~\ref{fig:ablation_fig} further shows that both variants recover largely overlapping sets of correct completions, confirming that the quality of the ranking stems primarily from the novel scoring mechanism itself.
The new metric Exact Match (EM) is representative of the actual generated output of the LLM. This metric matches Recall@1 when the generation is controlled. 

\emph{This important result demonstrates that \namemethod{} can be applied effectively even in scenarios where model generation cannot be controlled or constrained.}

\begin{table}[t]
\centering
\caption{Ablation study: impact of constrained decoding on ranking performance}
\label{tab:ablation_constraint_fixed}
\resizebox{\columnwidth}{!}{\begin{tabular}{lc cccc cccc}
\toprule
& & \multicolumn{4}{c}{\textbf{StartingPoints}} & \multicolumn{4}{c}{\textbf{DotPrompts}} \\
\cmidrule(lr){3-6} \cmidrule(lr){7-10}
\textbf{Model} & \textbf{Const.} & \textbf{EM} & \textbf{MRR} & \textbf{R@1} & \textbf{R@5} &\textbf{EM} &\textbf{MRR} & \textbf{R@1} & \textbf{R@5} \\
\midrule
SmolLM2-135M
& \ding{51}  & 0.62 & 0.73 & 0.62 & 0.88  & 0.67 & 0.76 & 0.67 & 0.87 \\
& \ding{55}  & 0.56 & 0.72 & 0.61 & 0.87  & 0.62 & 0.75 & 0.66 & 0.87 \\
\midrule
CodeGen-350M
& \ding{51}  & 0.70 & 0.78 & 0.70 & 0.89 & 0.77 & 0.83 & 0.77 & 0.91 \\
& \ding{55}  & 0.64 & 0.77 & 0.69 & 0.89  & 0.72 & 0.83 & 0.76 & 0.91 \\
\midrule
SmolLM2-360M
& \ding{51}  & 0.69 & 0.78 & 0.69 & 0.90  & 0.75 & 0.82 & 0.75 & 0.90 \\
& \ding{55}  & 0.63 & 0.78 & 0.68 & 0.90  & 0.69 & 0.81 & 0.74 & 0.90 \\
\midrule
DeepSeek-Coder 1.3B
& \ding{51}  & 0.78 & 0.84 & 0.78 & 0.92 & 0.83 & 0.88 & 0.83 & 0.94 \\
& \ding{55}  & 0.71 & 0.84 & 0.77 & 0.92  & 0.78 & 0.87 & 0.82 & 0.94 \\
\bottomrule
\end{tabular}}
\end{table}

\begin{figure}
  \centering
  \includegraphics[width=\columnwidth]{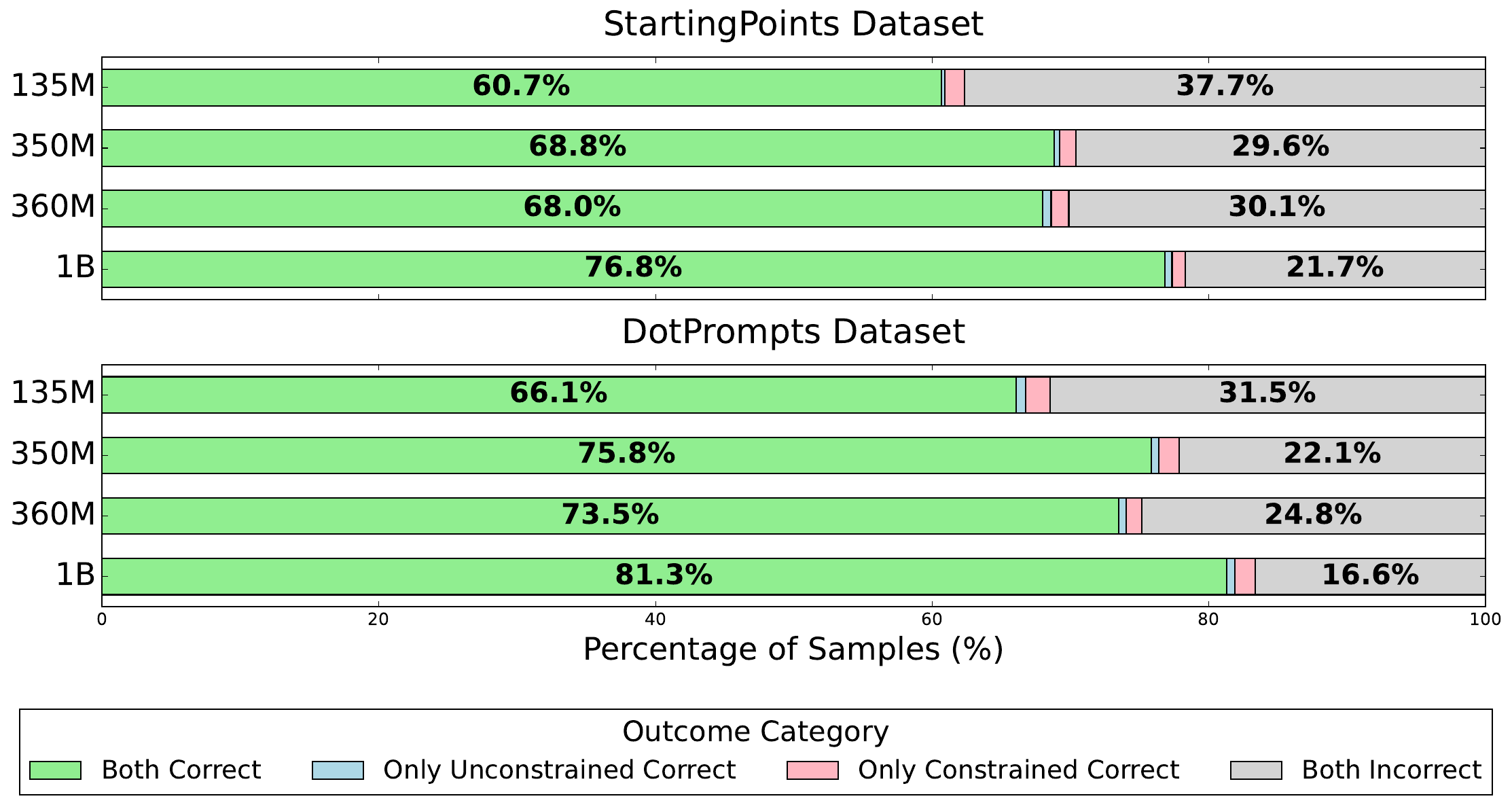}
  \caption{Recall@1 comparison of model outputs w/ and w/o constrained decoding.}
  \label{fig:ablation_fig}
\end{figure}


\section{Discussion}




Our results (Section~\ref{sec:rq1}) demonstrate that \namemethod{} consistently outperforms both traditional IDE ranking systems and common LLM baselines, matching the performance of computationally expensive reranking techniques while achieving up to a 30× speedup, across all tested model sizes and datasets. In \textit{DotPrompts}, our method achieves up to +16 MRR and +8 Recall@5 points over the best commercial IDE performance evem with the smallest model. Similar gains are observed in \textit{StartingPoints}, where completions target project-local identifiers, a setting where many neural systems degrade sharply due to lack of context.

Even under challenging conditions (See Table~\ref{tab:first_time_identifiers}), where the correct identifier is unseen in the prefix, \namemethod{} maintains high MRR and Recall@5. This indicates that our approach can still effectively distinguish semantically plausible candidates. This contrasts sharply with beam search baselines, which flatten in performance as beam width increases and offer only marginal ranking benefits at significant computational cost.

The efficiency metrics reinforce the practicality of our approach. \namemethod{} reduce minimal latency compared to greedy decoding, often outperforming beam-based strategies in both runtime and accuracy. With most completions resolved in one or two forward passes, and early termination occurring in over 75\% of cases, our method remains compatible with real-time usage in development environments.

\emph{In short, our method delivers substantial real-world value: better completions, faster inference, and smooth integration into existing IDE workflows. This sets a new baseline for what small models can achieve in interactive coding tools.}

\subsection{Differences between Java and Python Datasets}
The structure of identifiers differs substantially between Java and Python, shaping how completions unfold during decoding. Java identifiers (\textit{DotPrompts}) commonly follow \emph{camelCase}, producing compact tokenizations with fewer sub-word units. In contrast, Python identifiers (\textit{StartingPoints}) adopt \emph{snake\_case}, resulting in longer token sequences. This is reflected in the average ground-truth token length: 3.07 for Java versus 4.47 for Python (See \ref{tab:dataset-key-stats-inference}).
These differences impact decoding behavior. In \emph{DotPrompts}, only 35\% of completions are resolved in a single forward pass, yet 84\% are completed within two, indicating that most Java identifiers are fully predicted with minimal decoding overhead. In \emph{StartingPoints}, 57\% complete in one step, but gains taper off with additional passes (68\% within two). This suggests greater early predictability but more dispersed token structures.
These patterns confirm that identifier conventions affect decoding efficiency. Despite longer ground truth lengths, the token efficiency ratio remains high across both datasets, showing that \emph{\namemethod{}} accurately prioritizes relevant completions with fewer steps than naive greedy generation.

\subsection{Limitations and Future Directions}
We designed our approach with latency-aware constraints, although we do not claim real-time performance guarantees under a fully optimized production environment. All experiments were conducted using standard transformer implementations via HuggingFace~\cite{wolf_huggingfaces_2020}, without quantization or low-level runtime optimizations. As such, we do not report full end-to-end latency figures typical of production-grade systems. Nonetheless, our results show that the method achieves low-latency scoring even in this unoptimized setting. This suggests a strong potential for further improvements through integration with optimized inference runtimes and quantized models. We leave the implementation and evaluation of such a deployment to future work.

Another limitation of our current ranking strategy is the lack of explicit mechanisms to encourage diversity among top-ranked completions. In cases where multiple valid suggestions share the same prefix or semantic root, and receive identical scores from our decoding-based method, the resulting ranked list may lack variation. One possible solution is to filter out low-confidence tokens during traversal, allowing the model to focus only on the most probable continuations and reduce noise in score accumulation. Additionally, lightweight heuristics could be introduced to promote diversity, such as limiting the number of completions that share the same initial token. In ambiguous situations where several candidates remain indistinguishable by score, the system could fall back on existing ML-based ranking components already deployed in IDEs to refine or reorder the final top-k list. 

A promising direction for future work is to evaluate the performance of \namemethod{} within real-world IDEs to assess its impact on developer productivity and interaction patterns in practical coding scenarios. This evaluation was beyond the scope of the current work, which focuses on model design and controlled benchmarking.
\section{Related Work}
\label{sec:related_work}
\smartparagraph{LLM-Based Completion Ranking.}
Proposed LLM-based ranking systems commonly rely on \emph{beam search} to generate multiple candidate completions, followed by a separate learned model to re-rank them. The ML-enhanced code completion system from Google Research~\cite{tabachnyk_ml-enhanced_nodate} and the method by Li et al.~\cite{li_toward_2021} follow this architecture, treating the LLM purely as a generator. These systems ignore the structure of the code and perform ranking only after generation, resulting in unnecessary decoding overhead and limited semantic precision, making those approach sub-optimal in a limited resource environment. \textsc{FIRST}~\cite{reddy_first_2024} re-frames completion ranking as a prompt-based classification task, injecting the full list of completions into the prompt and requiring the LLM to select the correct one. Although this strategy shows promise in controlled settings, it \emph{requires general-purpose NLP reasoning and fine-tuning} on the task, making it incompatible with standard code-specific models that are typically deployed. Moreover, supporting this strategy would require shipping and maintaining a second model dedicated to ranking, adding memory and compute overhead that is impractical in constrained environments. In contrast, \textbf{our method is designed to integrate directly with existing completion engines}. \namemethod{} utilizes the same context as the generation model, allowing computational reuse, and is expressly model-agnostic, requiring no additional fine-tuning or separate models.
 


\smartparagraph{ML-based Ranking.} In this work, we focused on current implementations of ML-based ranking models used in the most popular IDEs on the market, using them as baselines to estimate our approach's impact.
However, other equally valid methodologies have also been explored. For instance, Svyatkovskiy et al.~\cite{10.1145/3292500.3330699} explored the use of LSTM-based models for capturing the semantic information embedded in code structure. Similarly, Asaduzzaman et al.~\cite{10.1109/ICSME.2014.29} proposed a context-sensitive technique that ranks suggestions based on similarity to previous usage examples. In contrast, our approach relies on attention mechanisms to extract relevant context information, which enables the model to dynamically focus on the most informative elements for each prediction.

\smartparagraph{Monitor-Guided Decoding} (MGD)~\cite{agrawal_monitor-guided_2023} proposes a method for injecting static analysis constraints into the LLM decoding process. BGD is a method for controlled code generation; this fundamental difference in objectives makes a direct quantitative comparison infeasible. Their implementation relies heavily on repeated conversions between strings and token IDs to maintain consistency with the model-generated tokens and completion strings. This introduces significant complexity and runtime overhead, making the method less suitable for latency-sensitive tasks like code completion ranking.

\smartparagraph{Tree-Based Representations.}
Kim et al.~\cite{kim_code_2021} propose a transformer-based approach for code completion that explicitly incorporates syntactic structure by feeding ASTs into the model. While their method improves precision for structured code, such as known API calls, it remains limited in handling out-of-vocabulary identifiers, as the AST vocabulary is predefined and static. In contrast, our approach operates at the token level and remains compatible with open vocabularies, enabling completion of both seen and unseen identifiers.



\section{Conclusions}
Recent findings show that 81\% of developers still rely on token-level completion, far exceeding 32\% who use statement-level suggestions~\cite{wang2023practitioners}. This underscores a critical point: code generation is only one aspect of modern programming assistance, while traditional token-level suggestions remain central to developer productivity. 
We introduced \namemethod{}, a lightweight decoding-time strategy that enhances code completion by leveraging language model probabilities to more effectively rank token-level suggestions such as identifiers and APIs.
Unlike conventional IDE systems that rely on hand-crafted heuristics or manually engineered features~\cite{bibaev_all_2022}, \namemethod{} integrates seamlessly with existing language models for code generation. It requires no prompt engineering, no model retraining, and no additional inference passes. By gathering fine-grained probabilities over a structured prefix tree during a single greedy decoding step, \namemethod{} produces accurate rankings at low computational cost, effectively extracting the model’s full potential in just a few forward passes.
Beyond its ranking accuracy, \namemethod{} meets the strict latency requirements of interactive environments. It achieves response times faster than standard greedy decoding and matches the quality of expensive reranking baselines. This balance of speed and precision makes it highly suitable for deployment in real-world IDEs.
Overall, our approach moves beyond isolated code generation and provides a lightweight, model-agnostic solution that bridges the gap between LLMs' capabilities and the low-latency demands of local IDEs tools.

\bibliographystyle{IEEEtran}
\bibliography{main}

\end{document}